# A new scheduling algorithm for server farms load balancing

Ehsan Saboori, Shahriar Mohammadi
K.N Toosi University of Technology
Tehran, Iran
ehsansaboori@sina.kntu.ac.ir
mohammadi@kntu.ac.ir

Shafigh Parsazad
Ferdowsi University of Mashhad
Mashhad, Iran
parsazad@stu.um.ac.ir

*Abstract*— this paper describes a new scheduling algorithm to distribute jobs in server farm systems .The proposed algorithm overcomes the starvation caused by SRPT (Shortest Remaining Processing Time). This algorithm is used in process scheduling in operating system approach . The algorithm was developed to be used in dispatcher scheduling. This algorithm is non-preemptive discipline, similar to SRPT, in which the priority of each job depends on its estimated run time, and also the amount of time it has spent on waiting. Tasks in the servers are served in order of priority to optimize the system response time. The experiments show that the mean round around time is reduced in the server farm system.

*Keywords*-server farm; load balancing; scheduling algorithm; dispatcher; round around time

## I. INTRODUCTION

Long round around time is a very important problem in internet services. There is more and more evidence showing a high variability of task size distribution in computer loads. For example, files requested in Web servers and in UNIX fit a heavy-tailed distribution [5]. One way to figure out this problem is using server farms. Server farm is consisting of a collection of many computers also called host, server or node and front-end high-speed dispatcher. Each incoming request is immediately dispatched via the dispatcher to one of the computers. The main advantages of using server farms are price and flexibility, because many slow computers are cheaper than fast computers and it is easy to up or down your server capacity. One of the most important issues in server farms is "routing policy", also known as "task assignment policy". This is the algorithm/rule for determining how to assign jobs to hosts. On the Internet, companies whose web sites get a great deal of traffic usually use load balancing. For load balancing Web traffic, there are several approaches. For Web serving, one approach is to route each request in turn to a different server. In some approaches, the servers are distributed over different geographic locations.

For distributing incoming jobs on the several servers, dispatcher is used. Dispatcher objective distributes incoming job on the servers in an efficient way. Dispatcher use scheduling algorithms to be able to dispatch jobs in the best way. Therefore scheduling algorithms is very important to achieve better performance in dispatcher. This paper describes a novel and efficient scheduling algorithm to be used in dispatcher for better job distributing in order to increase server farms performance.

## II. WEB SERVER FARMS

A Web server farm refers to a Web site that uses two or more servers housed together in a single location to handle user requests. They use one host called site to provide a single interface for users [1]. Today busy web servers are required to service many clients simultaneously, sometimes up to tens of thousands of concurrent clients [8]. Large web-server farms consist of thousands of servers and may handle millions of HTTP requests per second. These sites are overwhelmed by the offered load and the Web service provider shave to deal with peak demands that are much higher than the average load supported by their site. While operators are just about to be able to collect detailed traffic statistics, the very detail and volume make them nearly impossible to analyse. As a result, performance prediction, monitoring and performance measurement are rendered increasingly complicated, one computer cannot be able to handle the requests, no matter how many disks are used in parallel. The slow response times and difficult navigation are the most common complaints of Internet users [9]. Research shows the need for fast response time. The response time should be around 8 seconds as the limit of people's ability to keep their attention focus while waiting [10]. To figure out this problem more Computers must be added. Server farms consist of two important parts. First part is called "Front-End "and second is called "Back-End". To deliver requests to the individual machines of a server farm, a device is needed to accept all incoming traffic and to assign arequest to a particular machine to handle [3]. These devices are called Front End Devices (FEDs), because they sit at the front of a server farm accepting all requests. A major problem now arises, if a similar workstation or PC is used for a FED, then the number of attached machines in the server farm is typical limited to ten or less [4]. The most important part of any Web farms is the part that assigns the incoming load. This entity is called "Dispatcher". In other words, the Dispatcher acts as a centralized global scheduler that receives the totality of the requests and routes them among the back-end servers of the Web farm [2]. The dispatcher may use different scheduling policies to assign the load to the nodes of Web servers. Each server machine in the web farm is uniquely identified with a private address to access. Figure 1 is illustrated the server farm model.

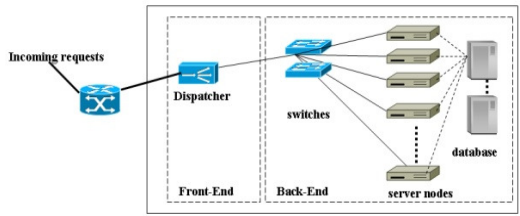

Figure 1. Server Farm model

### III. SCHEDULING ALGORITHMS

The problem with web server farms is that a load imbalance may occur, where some processors are idle with nothing to execute, while other processors are busy and have tasks in their ready queues. Any load imbalance will result in poorer task response times. Scheduling policies determine which requests in the queue are serviced at any point of time, how much time is spent on each, and what happened when a new request arrives. The goals of scheduling policies are to minimize the mean round around time of the request and to behave fairly to all requests [2].

Static algorithms are the fastest solutions. They do not have overloading decision making time. Static algorithms can potentially make poor assignment decisions, such as routing a request to a server node havinga long queue of waiting load while there are other almost idle nodes [5]. Dynamic algorithms have the potential to outperform static algorithms by using some state information to help dispatching decisions. On the other hand, they require mechanisms that collect, transmit and analyse state information there by incurring in overheads [1]. Round-Robin (RR), random (RAN) and Weighted Round-Robin (WRR) are dynamic algorithms.

#### A. FCFS (First-Come-First-Serve)

FCFS is a policy used to ensure fairness in a number of application domains such as scheduling [6] and Operating Systems [7]. This is a non-preemptive technique. A single queue of ready processes is maintained, and the dispatcher always picks the first one. This method does not emphasizes throughput, since long processes are allowed to monopolies CPU, For the same reason, the response time with FCFS can be high with respect to execution time, especially if there is a high variance in process execution times. It's fairly obvious that this method penalizes short processes following long ones, in appropriate for interactive systems; large fluctuations in average turn around time are possible, though no starvation is possible

#### B. SRPT (Shortest Remaining Processing Time)

Another policy that provable optimal mean round around time for all requests is SRPT "Shortest Remaining Processing Time". Shortest-Remaining-Processing-Time (SRPT) scheduling policy is an optimal algorithm for minimizing mean response time [11] and [12]. The SRPT scheduling policies on web servers [13], [14], and [15] used the job size, which is well known to the server, to refer to processing time (response time) of the job to implement SRPT for web servers to improve user-perceived performance. The job that has a least remaining process time will be served. There are two problems, this policy not fair. Jobs with large size may be waiting for a while and Dispatcher must know size of jobs beforehand.

#### C. Ps(Process Sharing)

In this policy capacity C of servers is equally shared between the incoming requests. This policy assures max-min fair allocation and easy to implement.

#### D. Multilevel Queue Scheduling

A multilevel queue scheduling algorithm partitions the ready queue into separate queues. For example, a common division is made between foreground (interactive) processes and background (batch) processes. Processes are permanently assigned to one queue. Each queue has its own scheduling algorithm. For example, foreground queue might be scheduled by a RR algorithm, while the background queue is scheduled by an FCFS algorithm. In addition, there must be scheduling between the queues. This is commonly a fixed-priority preemptive scheduling. For example, the foreground queue may have absolute priority over the background queue. Another possibility is to time slice between the queues. Each queue gets a certain portion of the CPU time, which it can then schedule among the various processes in its queue. For instance, in the foreground-background queue example, the foreground queue can be given 80 percent of the CPU time for RR among its processes, while the background queue receives 20 percent of the CPU to give its processes in a FCFS manner.

#### D. Round Robin

The round-robin scheduling algorithm sends each incoming request to the next server in its list. Thus in a three server cluster (servers A, B and C) request 1 would go to server A, request 2 would go to server B, request 3 would go to server C, and request 4 would go to server A, thus completing the cycling or "round-robin" of servers. It treats all real servers as equals regardless of the number of incoming connections or response time each server is experiencing.

#### E. Weighted Round Robin

The weighted round-robin scheduling is designed to better handle servers with different processing capacities. Each server can be assigned a weight, an integer value that indicates the processing capacity. Servers with higher weights receive new connections first than those with fewer weights, and servers with higher weights get more connections than those with fewer weights and servers with equal weights get equal connections. For example, the real servers, A, B and C, have the weights, 4, 3, 2 respectively, a good scheduling sequence will be AABABCABC in a scheduling period (mod sum Wi).

In the implementation of the weighted round-robin scheduling, a scheduling sequence will be generated according to the server weights after the rules of Virtual Server are modified. The network connections are directed to the different real servers based on the scheduling sequence in a round-robin manner. The weighted round-robin scheduling is better than the round-robin scheduling, when the processing capacity of real servers are different. However, it may lead to dynamic load imbalance among the real servers if the load of the requests varies highly. In short, there is the possibility that a majority of requests requiring large responses may be directed to the same real server. Actually, the round-robin scheduling is a special instance of the weighted round-robin scheduling, in which all the weights are equal.

## IV. THE PROPOSED ALGORITHM

In this paper, consider a new scheduling algorithm in server farms called HRRN. This algorithm used in process scheduling in operating system approach. This algorithm developed to use in dispatcher scheduling. This algorithm is discussed in this paper and shown how to evaluate it. HRRN is a variation on SRPT to solve a problem whereby long tasks may never get CPU time. If you imagine a system running with the SRPT algorithm that has a steady stream of processes coming in, it may be the case that a really long process never runs because there is always a shorter task waiting for the CPU. The HRRN algorithm fixes this by adjusting the priority of processes which are waiting to be run. If a process which will take a long time waits around for a while as a bunch of shorter processes come and go the system shortens the time that the scheduler thinks the long process will take. This of course doesn't make the long process complete any faster, but it does make it more likely to be scheduled. This repeats as long as the big job is waiting. Eventually, as the scheduler thinks the long job will be shorter and shorter, it is guaranteed to get the CPU. Just how long this will take depends on how the system is designed. Figure 2 is illustrate how can calculate priority of jobs in this algorithm.

$$Priority = \frac{waiting\_time + estimated\_run\_time}{estimated\_run\_time} = 1 + \frac{waiting\_time}{estimated\_run\_time}$$

Figure 2. Priority in HRRN scheduling algorithm

## V. THE SIMULATOR

The simulator was implemented using the C programming language on an IBM PC with Pentium 4 processors and 1GB of RAM, running Microsoft Windows XP. The simulator has two primary components: The job generator and the job dispatcher. The job generator is responsible for generating jobs randomly. Jobs are sequentially saved in a file in order to be served in servers. The job dispatcher is responsible for selecting jobs in the correct order to reduce the average round around time. Finally, round around time of each job is saved in an output file and charted in Microsoft Excel.

## VI. SCENARIO

At first, jobs are generated randomly with random served time, and then saved in a file. The dispatcher reads jobs from the file. Then the Dispatcher selects the best jobs to reduce the average round around time according to the aforementioned algorithm. The FCFS, SRPT and HRRN algorithms are implemented in the simulator. In this case, 100 jobs were generated and the dispatcher distributed the jobs on different number of servers. In this simulation, the Average round around time was calculated for each algorithm.

The SRPT has the best performance load balancing and average round around time, but starvation may be occur when it is used. Therefore this algorithm is not practical and is mainly used in comparison to other algorithms, i.e. the closet the average round around time of an algorithm to that one of SRPT, the better. In this simulation, different number of servers was used to analyse its impact on the average round around time. Algorithms act differently when the number of servers changed. When server count is 1, HRRN behaves the same as SRPT algorithm. When there are a small number of servers, HRRN behaves approximately similar to SRPT algorithm. A large number of servers will improve the performance of HRRN and SRPT algorithms over FCFS algorithms. Figure 3 illustrates the scenario of simulation for evaluation and comparison of the algorithms.

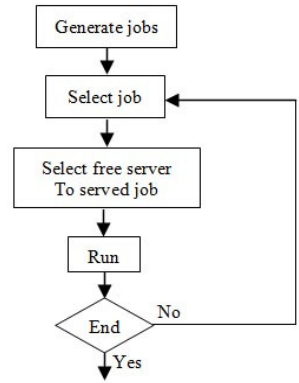

Figure 3. Scenario of simulation

## VII. COMPARISON

To analyse and compare the scheduling algorithms presented in the last section, the load balancer simulator was used. At first, 100 jobs are generated randomly with random estimated times. Then the simulator dispatches incoming jobs on servers and calculates the served time for each job. The mean round around time is also computed all jobs. In order to study the dependency of algorithms on the number of servers in server farms, the simulator was configured for different number of servers with the same defined jobs. The results are different based on the number of servers used. This means that the number of servers impacts on the performance of the algorithms. The results are present in Table1. This table illustrates the mean round around time for each algorithm, based on the number of servers used.

TABLE 1. The simulation results

| Number of servers | 25 | 15 | 6 | 1 |
|---|---|---|---|---|
| FCFS | 2342 ms | 3657 ms | 8588 ms | 49740 ms |
| SRPT | 1880 ms | 2763 ms | 6116 ms | 34174 ms |
| HRRN | 2034 ms | 2976 ms | 6383 ms | 34179 ms |

As illustrated in fig 4, when there is only one server used the result of HRRN and SRPT algorithms are the same. The vertical axis indicates the number of jobs and horizontal axis indicates the mean round around time in milliseconds.

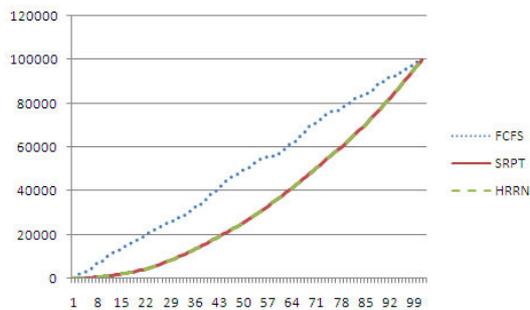

Figure 4. Results for one server configuration

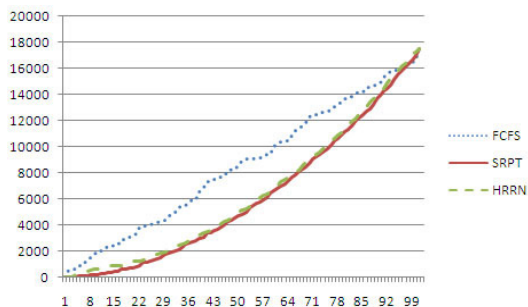

Figure 5. Results for six server configuration

This graph shows that HRRN and SRPT algorithms act similarly. According to Table 1, the number of servers impacts on the server farm performance. As the number of servers decreases, the functionality of HRRN algorithm becomes more similar to that of SRPT. As the number of servers increases, the difference between HRRN round around time and SRPT round around time grows.

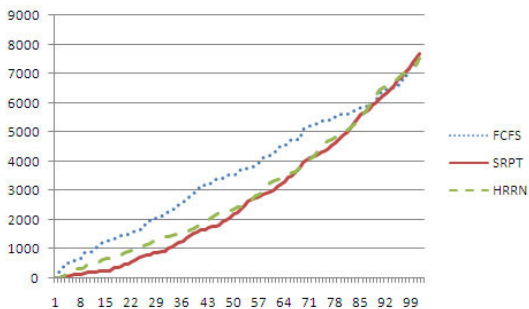

Figure 6. Results for 15 server configuration

Increasing the number of servers has a small impact on this difference .This fact is illustrated in Fig 6 and 7.

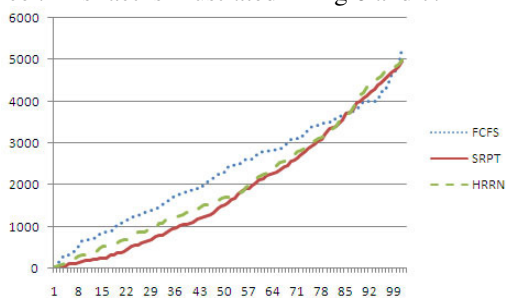

Figure 7. Results for 25 server configuration

VIII. CONCLUSION

The shortest remaining process time (SRPT) algorithm has the best performance among the scheduling algorithms but suffers from starvation. First-Come-First-Served (FCFS) algorithm is fairness but the mean round around time is too high. Any scheduling algorithm that is more similar to SRPT algorithm has better a performance. The proposed algorithm described in this paper uses HRRN algorithm in server farms dispatcher to distribute incoming jobs on the servers. The advantage of using this algorithm is better performance. The mean round around time of HRRN algorithm is more similar to that of SRPT. The most advantage of HRRN algorithm is overcoming starvation. This algorithm does not suffer from any starvation and has the same performance as SRPT does; therefore using the HRRN algorithm in the server farms dispatcher can increase the performance.